\documentclass[12pt,preprint]{aastex}

 \usepackage{epstopdf}
 \slugcomment{Accepted Version 1}

\newcommand\pv{\mbox{$p_{V}$}}
\newcommand\pIR{\mbox{$p_{IR}$}}
\newcommand\irfactor{\mbox{$p_{IR}/p_{V}$}}

\begin{document}

 \DeclareGraphicsExtensions{.pdf,.gif,.jpg}

 \title{Physical Parameters of Asteroids Estimated from the WISE 3 Band Data and NEOWISE Post-Cryogenic Survey}
\author{A. Mainzer\altaffilmark{1}, T. Grav\altaffilmark{2}, J. Masiero\altaffilmark{1}, J. Bauer\altaffilmark{1,3}, R. M. Cutri\altaffilmark{3}, R. S. McMillan\altaffilmark{4}, C. R. Nugent\altaffilmark{5}, D. Tholen\altaffilmark{6}, R. Walker\altaffilmark{7}, E. L. Wright\altaffilmark{8}}

 \altaffiltext{1}{Jet Propulsion Laboratory, California Institute of Technology, Pasadena, CA 91109 USA}
\altaffiltext{2}{Planetary Science Institute, Tucson, AZ USA}
\altaffiltext{3}{Infrared Processing and Analysis Center, California Institute of Technology, Pasadena, CA 91125, USA}
\altaffiltext{4}{Lunar and Planetary Laboratory, University of Arizona, 1629 East University Blvd., Tucson, AZ 85721-0092, USA}
\altaffiltext{5}{Department of Earth and Space Sciences, UCLA, 595 Charles Young Drive East, Box 951567, Los Angeles, CA 90095-1567 USA}
\altaffiltext{6}{Institute for Astronomy, University of Hawaii, 2680 Woodlawn Drive, Honolulu, HI USA}
\altaffiltext{7}{Monterey Institute for Research in Astronomy, Monterey, CA USA}
\altaffiltext{8}{Department of Physics and Astronomy, UCLA, PO Box 91547, Los Angeles, CA 90095-1547 USA}

 \email{amainzer@jpl.nasa.gov}

 \begin{abstract}
Enhancements to the science data processing pipeline of NASA's Wide-field Infrared Explorer (WISE) mission, collectively known as NEOWISE, resulted in the detection of $>$158,000 minor planets in four infrared wavelengths during the fully cryogenic portion of the mission. Following the depletion of its cryogen, NASA's Planetary Science Directorate funded a four month extension to complete the survey of the inner edge of the Main Asteroid Belt and to detect and discover near-Earth objects (NEOs).  This extended survey phase, known as the NEOWISE Post-Cryogenic Survey, resulted in the detection of $\sim$6500 large Main Belt asteroids and 88 NEOs in its 3.4 and 4.6 $\mu$m channels.  During the Post-Cryogenic Survey, NEOWISE discovered and detected a number of asteroids co-orbital with the Earth and Mars, including the first known Earth Trojan. We present preliminary thermal fits for these and other NEOs detected during the 3-Band Cryogenic and Post-Cryogenic Surveys.

 \end{abstract}

 \section{Introduction}

NASA's \emph{Wide-field Infrared Survey Explorer} mission \citep[WISE;][]{Wright.2010a, Cutri.2012a} launched on 14 December 2009 and operated until placed into hibernation on 17 February 2011.  WISE surveyed the entire sky near 90$^\circ$ solar elongation in four infrared wavelengths: 3.4, 4.6, 12 and 22 $\mu$m (denoted $W1$, $W2$, $W3$, and $W4$ respectively).  This scan pattern resulted in an average of 12 exposures on the ecliptic, rising to hundreds of exposures at the ecliptic poles with 6.5 arcsecond spatial resolution at 12 $\mu$m. Cooling for all four detectors was provided by dual solid hydrogen tanks.  Survey operations began on 7 January, 2010, and the first pass on the entire sky was completed six months later.  Coverage of the solar system was incomplete at that time owing to the long synodic periods of near-Earth objects (NEOs) and many Main Belt asteroids (MBAs). The outer hydrogen tank was exhausted on 5 August, 2010, resulting in the loss of the 22 $\mu$m channel that day.  As the remaining hydrogen ice in the inner tank sublimated, both the detectors and the telescope temperature rose.  During this period, the 12 $\mu$m channel continued to operate (albeit with reduced sensitivity) until 29 September, 2010.  The inner tank's hydrogen supply was then finally exhausted, and the telescope temperature rose, resulting in the loss of this channel.  

Before launch, the WISE baseline data processing pipeline was enhanced with the WISE Moving Object Pipeline System (WMOPS) to enable the independent discovery of new minor planets in near real-time and archive individual exposures.  These augmentations to the WISE pipeline, collectively known as ``NEOWISE," resulted in the discovery of $\sim$34,000 new asteroids, including 135 new NEOs.  The survey has reported observations of $>$158,000 minor planets \citep{Mainzer.2011a}.  

Beginning 1 October, 2010, NASA's Planetary Science Directorate funded a four month extension, known as the NEOWISE Post-Cryogenic Survey, to search for new asteroids using WMOPS and to fill in the gap in coverage of the inner Main Asteroid Belt \citep[see Figure 1 of ][]{Mainzer.2011a}.  The focal plane assemblies, optics and telescope temperature warmed to 73.5 K, cold enough to permit observations in the two shortest wavelengths.  As discussed in \citet{Cutri.2012a}, the $W1$ and $W2$ arrays operated with minimal performance degradation.  Data collection was halted on 1 February, 2011.

To date, the NEOWISE Post-Cryogenic Survey data have been processed using a first-pass version of the reduction pipeline.  The image data were calibrated using early versions of dark and flat field images and bad pixel masks that were largely derived from the fully cryogenic (FC) mission phase.  The current processing version does not benefit from the improvements implemented in the second-pass processing that have been applied to the WISE All-Sky and 3-Band Cryogenic Data Release products.  Furthermore, the precise calibrations needed to compensate optimally for the effects of the increasing temperature were not applied to the Post-Cryogenic Survey data.  For example, the $W1$ and $W2$ system throughput is known to have changed with time as the focal plane arrays and telescope warmed.  Comparison of photometry of inertial calibration sources (primarily stars) between the four band data and the Post-Cryogenic Survey data indicates that objects may appear up to 2\% brighter in $W1$ and up to 11\% brighter in $W2$ during different times throughout the Post-Cryogenic Survey \citep{Cutri.2012a}. 

The NEOWISE Post-Cryogenic Survey preliminary data were delivered to NASA's public Infrared Science Archive on 31 July, 2012.  These preliminary release data do not have the same degree of quality assurance as the final release products delivered for the fully cryogenic WISE All-Sky Data Release. Consequently, users are strongly advised to consult the cautionary notes in Section VIII.1.d of the Post-Cryo Preliminary Data Release Explanatory Supplement \citep[][\it{http://wise2.ipac.caltech.edu/docs/release/allsky/expsup/}]{Cutri.2012a}.  The NEOWISE project is now reprocessing the Post-Cryogenic Survey data with a second-pass version of the pipeline, using the much improved calibrations optimized for the warm telescope.

Here, we characterize the performance of the preliminary version of the Post-Cryogenic Survey data with regard to small bodies.  To date, $\sim$6500 objects were detected during the NEOWISE Post-Cryogenic Survey, including 88 NEOs.   NEOWISE discovered $\sim$1000 new asteroids during the Post-Cryogenic Survey, including 12 new NEOs.  We present preliminary thermal model results for NEOs detected during the 3-Band Cryogenic and Post-Cryogenic Survey phases.  NEOWISE discovered two unusual NEOs during the Post-Cryogenic Survey: 2010 TK$_{7}$, the first known Earth Trojan \citep{Connors.2011a}, and 2010 SO$_{16}$, an NEO in a so-called ``horseshoe" orbit \citep{Christou.2011a}. The fits for these and other objects are given.    

\section{Methods}
\subsection{Accuracy of Thermal Models Applied to Post-Cryogenic Data}
We consider the accuracy of thermal model fits for small body measurements using the NEOWISE Post-Cryogenic Survey data in two ways.  First, we compare fit results from data collected during the FC mission for a number of objects with well-known diameters to fits performed using only the $W1$ and $W2$ bands from the same FC dataset (i.e. drop bands $W3$ and $W4$).  Next, we compare fits performed using the FC data for $\sim$430 NEOs to fits for the same objects using only their $W1$ and $W2$ data (again dropping $W3/W4$).  

At 3.4 and 4.6 $\mu$m, asteroid fluxes are a mix of reflected sunlight and thermal emission.  While flux at 12 $\mu$m is dominated by thermal emission for most NEOs and MBAs, which is only a weak function of visible albedo (\pv), the fluxes at shorter wavelengths are strong functions of \pv.  We make the simplifying assumption that $p_{IR} = p_{3.4 \mu m} = p_{4.6 \mu m}$.  In \citet{Mainzer.2011b}, we showed that infrared albedos are correlated with taxonomic classification.  Asteroids that are red-sloped (such as T and D types) from visible to near-infrared wavelengths tend to have the higher ratios of \irfactor, and asteroids with blue to flat slopes (such as C and B types) tend to have lower \irfactor\ ratios.    

We apply the Near-Earth Asteroid Thermal Model \citep[NEATM;][]{Harris.1998a} to obtain estimates of the asteroids' physical parameters.  This model uses a beaming parameter, $\eta$, to account for the tendency of surface features such as craters to ``beam" radiation back to the observer at low phase angles.  When two or more thermally-dominated bands are available, $\eta$ can be estimated, and this was done for more than a hundred thousand asteroids observed by WISE during the FC survey phase \citep{Mainzer.2011c, Masiero.2011a, Grav.2011a, Grav.2012a}.  These works implemented the NEATM code in such as way as to determine the free parameters $D$, \pv, \pIR, and $\eta$ by using a least-squares minimization to the four wavelengths and their errors, along with absolute visible magnitude $H$ and its assumed error (taken to be $\pm$0.3 mag in most cases), and slope parameter $G$ and its error ($\pm$0.1 mag).

However, for objects detected only during the Post-Cryogenic Survey, it is not usually possible to fit for $\eta$ for most inner solar system objects, because only the $W2$ band is thermally dominated. For asteroids at sufficiently large heliocentric distances, even $W2$ is primarily dominated by reflected sunlight.  Hence, for most inner Main Belt asteroids and NEOs observed in only the Post-Cryogenic Survey, we must assume a value for $\eta$, along with an estimate of its variance.  For NEOs, we adopted $\eta=1.4\pm0.5$, following \citet{Mainzer.2011c}, while for MBAs, we set $\eta=1.0\pm0.25$ based upon the average values shown in \citet{Masiero.2011a}.  Following \citet{Mainzer.2011d}, errors in derived physical parameters such as diameter and \pv\ were computed using Monte Carlo trials which varied the WISE magnitudes, absolute visible magnitude $H$ according to their 1-$\sigma$ errors, along with any other assumed parameters such as $\eta$ or \pIR\ where necessary.  

When measurements in all four WISE wavelengths and an absolute visible magnitude $H$ are available, it is possible to fit for \pIR\ directly rather than compute it from \pv\ using an assumed value for the ratio \irfactor.  However, when there are insufficient measurements dominated by reflected sunlight available, it is no longer possible to fit \pIR.  In that case, an assumed value for \irfactor\ must be used.  We used \irfactor=1.6$\pm$1.0 for NEOs based \citet{Mainzer.2011d}, and \irfactor=1.3$\pm$0.5 for MBAs based on \citet{Masiero.2011a}.  

For objects detected by WISE at 12 and 22 $\mu$m, we have shown in \citet{Mainzer.2011d} that effective spherical diameters can generally be determined to within $\pm$10\%, and albedos to within $\pm$25\% of their values.  We assembled a collection of 112 asteroids from the literature with reliable diameters that were measured with non-radiometric methods (radar observations, stellar occultations, or spacecraft visits).  Here, we calculate thermal fits for these calibration objects using only bands $W1$ and $W2$.    

\begin{figure}
 \figurenum{1}
\includegraphics[width=6in]{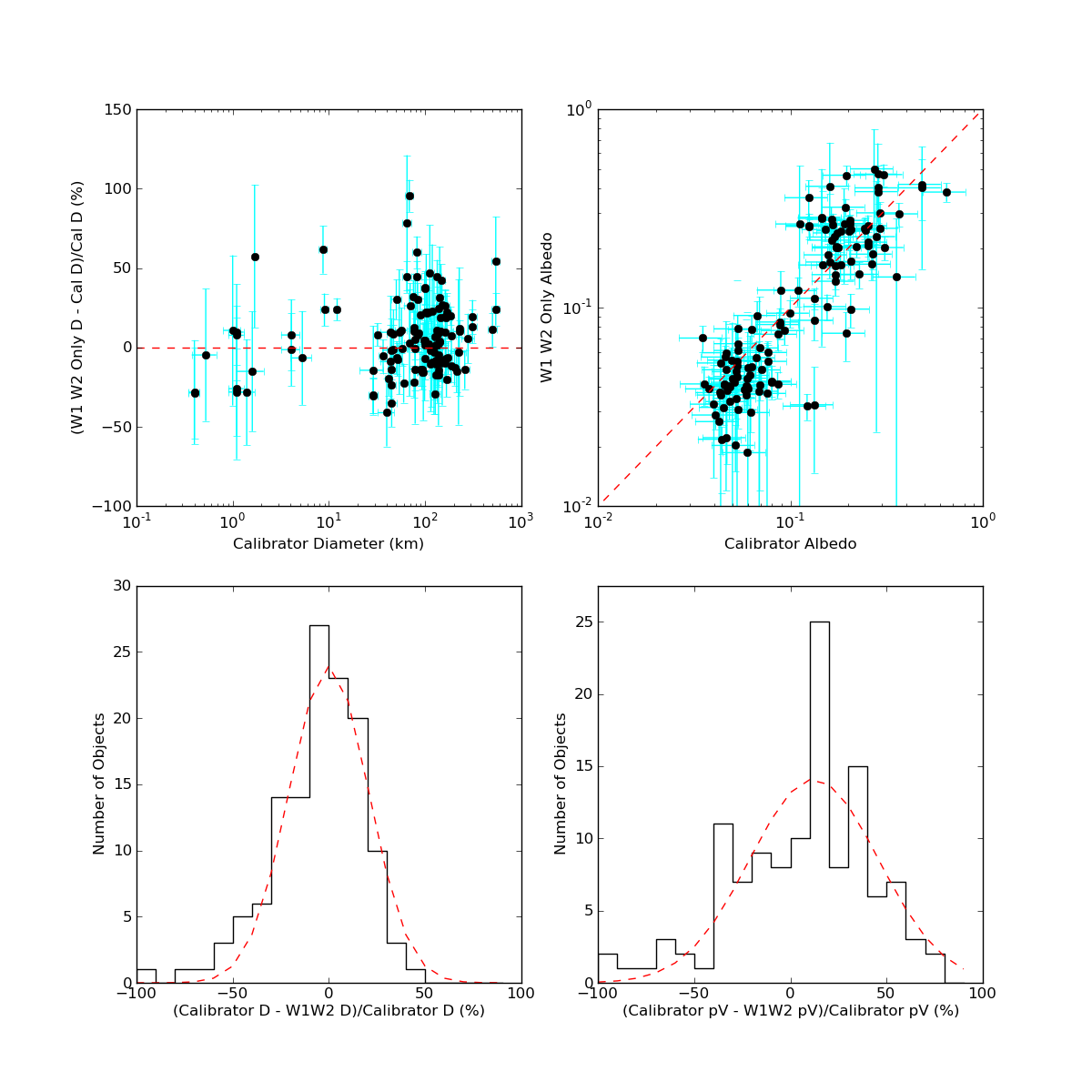}
\caption{\label{fig:w1w2_v_radar1}Asteroid diameters and albedos derived from observations during the FC WISE mission using only 3.4 and 4.6 $\mu$m are compared to diameters obtained from either radar, stellar occultations, or in situ spacecraft visits.  Gaussian fits to the fractional difference between diameters and albedos are shown as dashed red lines in the lower two panels.  The diameters derived from $W1/W2$ only reproduce the calibrator targets' diameters to within $\pm20\% (1-\sigma)$, and visible albedos to within $\pm40\% (1-\sigma)$.   }
\end{figure}

Figure \ref{fig:w1w2_v_radar1} shows the results of NEATM fits for the 112 calibrator objects from data from the FC survey phase, but using only $W1/W2$ measurements.  Gaussian fits to the resulting histogram of differences between $W1/W2$-only fits and the diameters taken from non-radiometric literature sources produces a 1-$\sigma$ diameter spread of $\sim$20\%, and a 1-$\sigma$ spread in albedos of $\sim$40\% of the objects' albedos.  The accuracy of the diameters and albedos derived using $W1/W2$ only is degraded by about a factor of two relative to fits derived using $W3/W4$.

Figure \ref{fig:w1w2_v_radar1} shows that fits for \pv\ for objects with lower \pv\ tend to be somewhat lower than those derived from radar, etc.    Similarly,  fits for \pv\ for higher albedo objects may be somewhat systematically too high.  As noted in \citet{Mainzer.2011b}, S-complex asteroids tend to have redder visible-to-near-infrared slopes, so therefore they tend to have higher values of \irfactor. As most of the objects shown in Figure \ref{fig:w1w2_v_radar1} are MBAs, the assumed value used for \irfactor\ of 1.3$\pm$0.5 is probably too low; \citet{Mainzer.2011b} found a median value of $\sim$1.6 for S-complex asteroids.  Moreover, \citet{Mainzer.2011b} found a median \irfactor\ of $\sim$1.0 for C-complex asteroids, as might be expected from their blue to flat visible-to-near-infrared slopes.  The ratio \irfactor\ seems to indicate that on average, for many objects, the trend of the visible to near-infrared slope continues out to 3-4 $\mu$m.  There are exceptions to these general trends; T- and D-type objects, which have lower \pv, have red to extremely red slopes and median \irfactor\ of 1.5 and 2.0, respectively.  The one-size-fits-all assumption of \irfactor=1.3$\pm$0.5 for MBAs and 1.6$\pm$1.0 for NEOs when performing thermal fits using $W1/W2$ only is perhaps overly simplistic. Nevertheless, given that taxonomic classification is not usually available for most objects, it is probably not safe to assume different values for \irfactor.  The errors are more properly accounted for by using an appropriately large error on the assumed value of \irfactor.      

\begin{figure}
 \figurenum{2}
\includegraphics[width=6in]{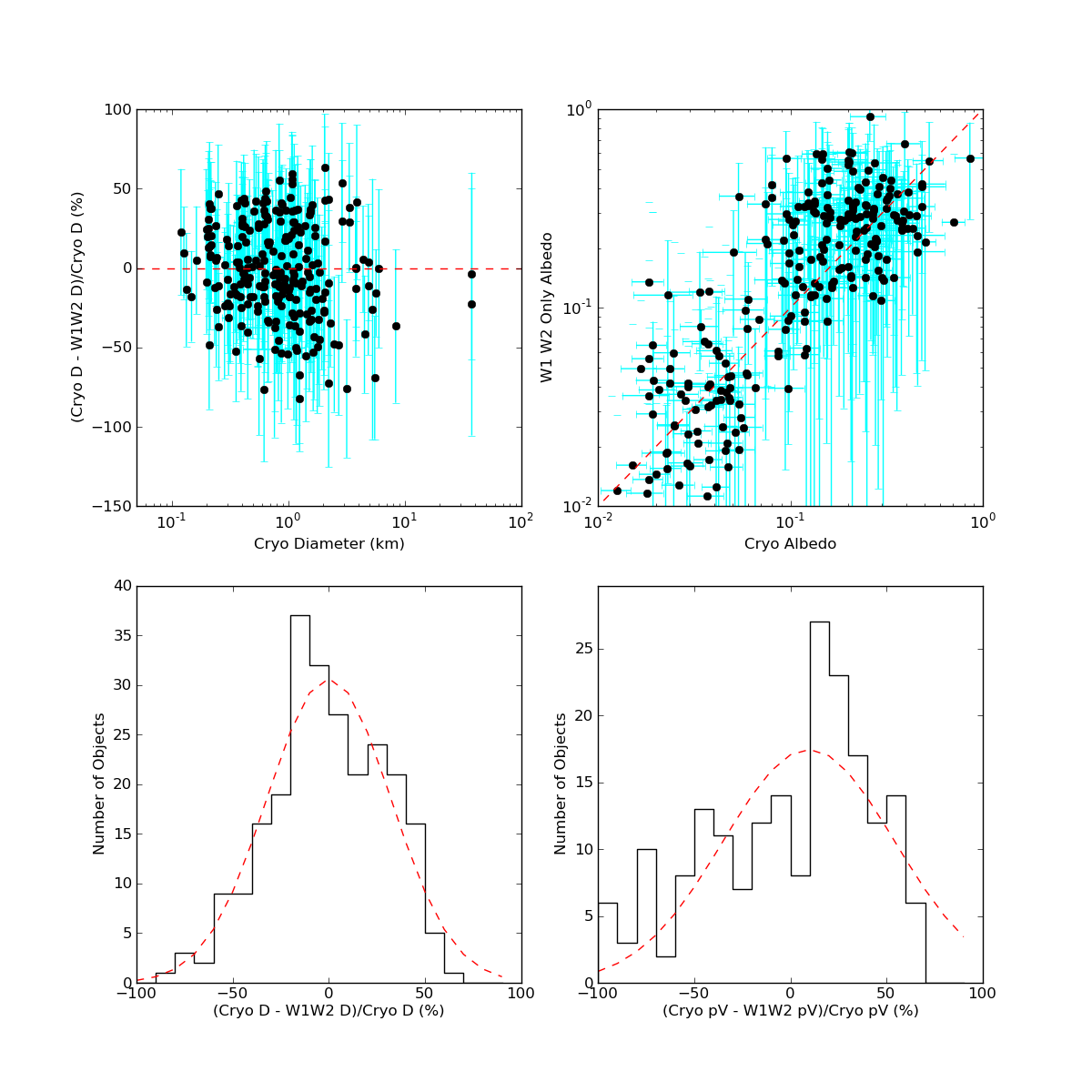}
\caption{\label{fig:w1w2_v_cryoneos1}Asteroid diameters and albedos derived from observations of NEOs during the FC survey using all four WISE bands (3.4, 4.6, 12 and 22 $\mu$m) are compared to thermal fits derived by using only 3.4 and 4.6 $\mu$m from the same dataset (i.e. disregarding 12 and 22 $\mu$m).  The red dashed lines represent Gaussian fits to the fractional differences between four band and two band thermal fits.  As with the objects shown in Figure \ref{fig:w1w2_v_radar1}, diameters computed using only bands $W1/W2$ match the four band fits to within $\sim \pm20\% (1-\sigma)$, and visible albedos to within $\sim \pm40\% (1-\sigma)$.  These errors are about a factor of two worse than fits derived including 12 and 22 $\mu$m.}
\end{figure}

We next compared thermal fit results derived for 429 NEOs observed during the FC survey phase with fits derived using their $W1/W2$ detections only (i.e. dropping 12 and 22 $\mu$m).  Of the 429 NEOs detected in the FC survey phase, 218 were detected in bands $W1/W2$ in addition to $W3/W4$.  By comparing the FC fits to fits for the same objects using only their $W1/W2$ fluxes, we can assess the difference in diameters and albedos (Figure \ref{fig:w1w2_v_cryoneos1}).  The $W1/W2$-only fits replicate the four band fits to within $\pm$20\% in diameter and $\pm$40\% in albedo.  While the mean offset between the $W1/W2$-only diameters and the four band diameters is zero, there is a skew in the histogram of differences between $W1/W2$-only albedos and four band albedos.  As noted with the calibration objects shown in Figure \ref{fig:w1w2_v_radar1}, when performing thermal fits using \pIR=1.6$\pm$1.0 (the median value derived from the FC survey phase for NEOs), NEOs with \pv$>\sim$0.1 tend to have somewhat larger albedos than their corresponding four band values.  Given that in \citet{Mainzer.2011b} it was shown that \pIR\ is most likely related to an object's visible/near-infrared slope, which is not known in most cases, we must use the median value of \pIR=1.6$\pm$1.0. Nevertheless, Figures \ref{fig:w1w2_v_radar1} and \ref{fig:w1w2_v_cryoneos1} show that the NEOWISE Post-Cryogenic Survey data produce reasonable estimates of an asteroid's effective spherical diameter and albedo. 

Table 1 gives the physical properties of 32 NEOs detected during both the 3-Band Cryogenic phase of the WISE mission and the 88 NEOs detected during the NEOWISE Post-Cryogenic Survey.  The median diameter of objects detected in the Post-Cryogenic survey phase was 0.87 km, vs. 0.76 km during the FC mission.

\subsection{Co-Orbital Asteroids}
Table 1 includes a number of unusual objects.  Several objects that are currently co-orbital with respect to the Earth were detected, including (3753) Cruithne.  Cruithne's orbit, when viewed from a reference frame corotating with Earth, follows a horseshoe-shaped pattern \citep{Wiegert.1998a, Christou.2000a}.  NEATM yields an effective spherical diameter of 2.1$\pm$0.21 km.  Its high albedo ($p_{V}=0.37\pm0.09$) and moderately high infrared albedo ratio (\pIR=1.13$\pm$0.35) are consistent with its spectral classification of Q \citep{Xu.1995a,Bus.2002a}.  It exhibits a large-amplitude lightcurve that appears consistent with the 27.4 hour rotational period derived by \citet{Erikson.2000a}.

NEOWISE discovered another object that was determined to be in a horseshoe orbit with respect to the Earth by \citet{Christou.2011a}, 2010 SO$_{16}$, during the Post-Cryogenic Survey.  It was detected in band $W2$, and a NEATM fit assuming the standard values of $\eta=1.4\pm0.5$ and \pIR=1.6$\pm$1.0 determined from \citet{Mainzer.2011c} yields an effective spherical diameter of 0.35$\pm$0.14 km with a visible albedo $p_{V}=.08\pm0.06$.  

The first known Earth Trojan asteroid, 2010 TK$_{7}$, was discovered by NEOWISE during the Post-Cryogenic Survey.  \citet{Connors.2011a} analyzed its orbit and determined it to be bound to the Earth's L4 Lagrange point.  A NEATM fit to the Post-Cryogenic Survey data shows the object to be 0.38$\pm$0.12 km in diameter with $p_{V}=.06\pm0.05$.  2010 TK$_{7}$ passed through the WISE field of view for $\sim$80 exposures during the fully cryogenic portion of the mission, but it was not detected even in a comoving coadded image of all of these frames.  A prediction of its magnitude using the physical properties determined from the Post-Cryogenic Survey indicates that it would have fallen well below the detection threshold even with $\sim$80 exposures.  

A quasi-satellite was also detected during the fully cryogenic portion of the mission in bands $W2$ and $W3$, (164207) 2004 GU$_{9}$.  It will likely trail the Earth for the next $\sim$1000 years, after which point it will most likely enter a horseshoe orbit \citep{Wajer.2010a}.  It is small ($D=0.17\pm0.02$ km) and bright (\pv=$0.19\pm0.05$).     

NEOWISE also detected a known Mars Trojan, (121514) 1999 UJ$_{7}$.   The object was observed in two epochs, first during the fully cryogenic portion of the mission in bands $W2$, $W3$, and $W4$ in March, 2010, then again in October, 2010 in bands $W1$ and $W2$ only.  A NEATM fit reveals its diameter to be large, about 2.5$\pm$0.2 km, with a low visible albedo ($p_{V}$=0.048$\pm$0.012), consistent with its X or T classification \citep{Rivkin.2003a}.  Its low albedo is significantly different from those of other Mars Trojans with measured albedos such as (5261) Eureka and (101429) 1998 VF$_{31}$ \citep[$p_{V}$=0.39 and 0.32 respectively;][]{Trilling.2007a}.  The difference in albedo and spectral classification may indicate a different origin from the other Mars Trojans.    

\section{Discussion}
\begin{figure}
\figurenum{3}
\includegraphics[width=6in]{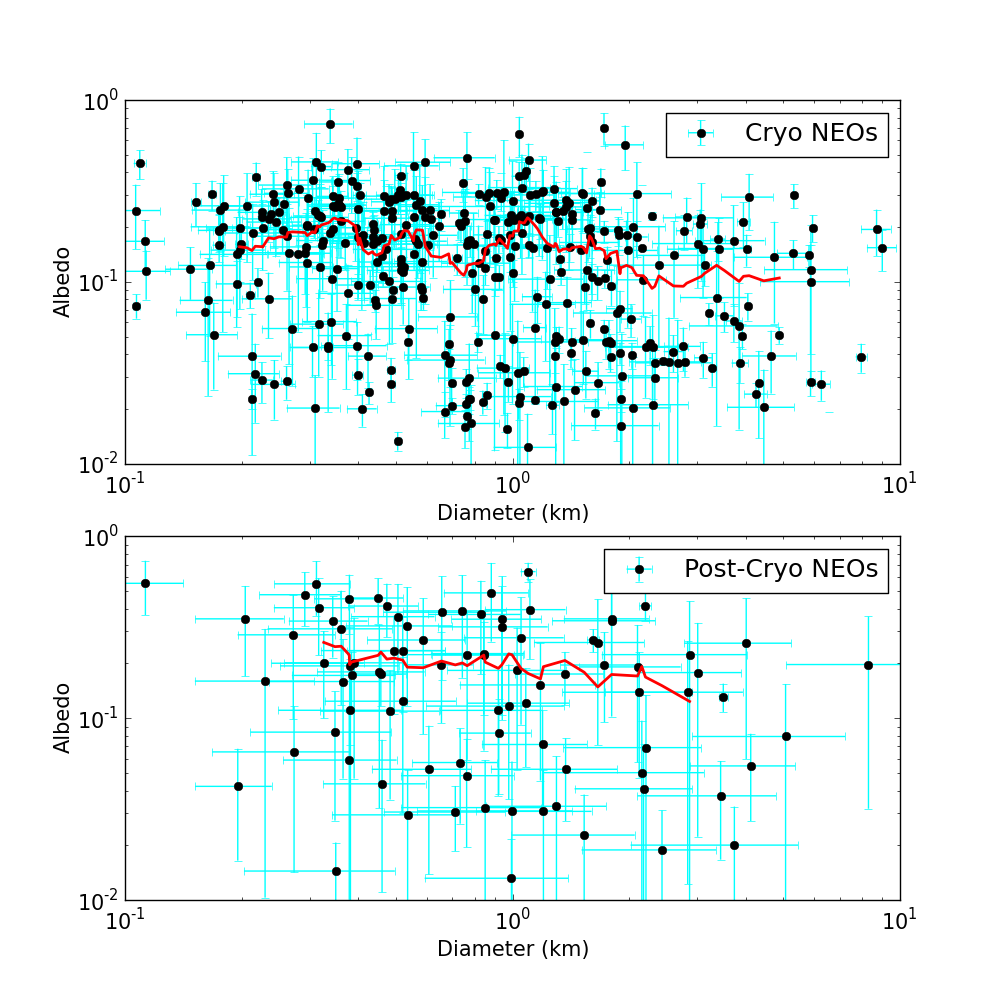}
\caption{\label{fig:albedo}There is no significant trend in albedo vs. diameter for the $\sim$420 NEOs detected by WMOPS during the fully cryogenic portion of the mission (top panel; running median shown in red), most of which were detected in the thermally-dominated 12 $\mu$m band $W3$.  There are an additional $\sim$10 NEO candidates that appeared on the Minor Planet Center's NEO Confirmation Page but did not receive follow-up, often because they were too faint to be detected by available assets.  These missing objects will tend to be small and dark.  Bottom panel: 88 NEOs detected at 4.6 $\mu$m during the Post-Cryogenic Survey have a non-negligible fraction of their flux that is produced by reflected sunlight, and the amount is proportional to their albedo, as indicated by the slight slope in the running median (red line).  Hence, somewhat fewer small, dark NEOs were detected during the Post-Cryogenic Survey relative to the fully cryogenic phase. }
\end{figure}

Most of the asteroids detected by WMOPS during the fully cryogenic phase of the WISE mission were identified using band $W3$, centered at 12 $\mu$m.  At this wavelength, the dominant source of flux for NEOs is thermal emission rather than reflected sunlight.  As thermal flux is only weakly dependent on $p_{V}$, band $W3$ is essentially equally sensitive to high and low albedo objects.  However, NEOs detected by WMOPS in band $W2$ during the Post-Cryogenic Survey were not selected in such an albedo-insensitive manner, given that flux in band $W2$ is usually a mix of reflected sunlight and thermal emission for most NEOs.  Darker, smaller NEOs are therefore somewhat less likely to have been detected.  Figure \ref{fig:albedo} shows the relationship between $p_{V}$ and diameter for NEOs observed during the fully cryogenic mission phase as well as during the Post-Cryogenic Survey.  As expected, somewhat fewer small, low albedo NEOs were detected by WMOPS during the fully cryogenic survey phase.

In order to extrapolate the sample of co-orbital near-Earth asteroids detected by NEOWISE to the larger population as was done for the fully cryogenic survey phase in \citet{Mainzer.2011c} and  \citet{Grav.2011a, Grav.2012a}, we must develop a synthetic population of objects with suitable orbital elements and physical properties.  This synthetic population will be combined with a model of the survey's sensitivity as a function of sky position and wavelength to produce estimates of the survey's biases with respect to orbital elements as well as diameter and albedo.  This analysis will be the subject of future work. 

\section{Conclusions}
The NEOWISE Post-Cryogenic Survey has resulted in the collection of 3.4 and 4.6 $\mu$m observations for nearly 100 NEOs to date.  Comparison with well-known calibrator objects shows that for objects with low amplitude lightcurves, good signal-to-noise measurements, and thermally-dominated fluxes in band $W2$, diameters can be computed to within $\pm$20\% and albedos to within $\pm$40\% of their values.  While the accuracy of the diameters and albedos computed from bands $W1$ and $W2$ only is lower when compared to the fully cryogenic mission data, it is nevertheless possible to learn useful physical information about these objects.  We have computed preliminary thermal models for the NEOs detected during the 3-Band Cryogenic and Post-Cryogenic Survey phases, including a number of objects that are co-orbital with the Earth and Mars.  Future work will include extrapolation of these detections to place constraints on the populations of these objects as well as strength and effect of non-gravitational forces on their orbital evolution.  

\section{Acknowledgments}

\acknowledgments{This publication makes use of data products from the Wide-field Infrared Survey Explorer, which is a joint project of the University of California, Los Angeles, and the Jet Propulsion Laboratory/California Institute of Technology, and NEOWISE, which is a project of the Jet Propulsion Laboratory/California Institute of Technology. WISE and NEOWISE are funded by the National Aeronautics and Space Administration.  We gratefully acknowledge the extraordinary services specific to NEOWISE contributed by the International Astronomical Union's Minor Planet Center, operated by the Harvard-Smithsonian Center for Astrophysics.  We thank the worldwide community of dedicated amateur and professional astronomers devoted to minor planet follow-up observations. We thank our referee for thoughtful comments.  This research has made use of the NASA/IPAC Infrared Science Archive, which is operated by the Jet Propulsion Laboratory, California Institute of Technology, under contract with the National Aeronautics and Space Administration.}

  \clearpage

 \clearpage
 
\begin{deluxetable}{lllllllllll}
\tabletypesize{\tiny}
\tablecolumns{11}
\tablecaption{NEATM results for the 32 NEOs detected by NEOWISE during the 3-Band Cryogenic Survey phase and the 88 NEOs detected during the NEOWISE Post-Cryogenic Survey. This table contains the preliminary thermal fit results based on the first-pass version of the WISE data processing as described in the text. The columns contain object name, $H$ magnitude, slope parameter $G$, diameter, \pv, beaming parameter $\eta$, $p_{IR}$, and number of observations in each of the four WISE bands.  The $1 -\sigma$ errors presented here were statistically generated using Monte Carlo modeling. WISE fluxes, absolute magnitude $H$, and $G$ were varied by their $1-\sigma$ error bars, as well as beaming ($\eta$) and $p_{IR}$ when these two parameters could not be fitted.  The quoted precision for each parameter follows the object with the most significant figures for the error on that value in the  table.  The statistical errors on diameter and \pv\ for each object in the table should be added in quadrature to the $\pm10\%$ and $\pm25\%$ systematic errors described in the text. Table 1 is published in its entirety in the electronic edition of the journal; a portion is shown here for guidance regarding its form and content.  }
\tablehead{\colhead{Object} & \colhead{$H$} &\colhead{$G$} & \colhead{$D$ (km)} & \colhead{\pv}  & \colhead{$\eta$} & \colhead{$p_{IR}$} & \colhead{N(W1)} & \colhead{N(W2)} & \colhead{N(W3)} & \colhead{N(W4)} }
\startdata
  01566 & 16.30 & 0.15 &    1.417 $\pm$    0.123 &    0.199 $\pm$    0.110 &    1.403 $\pm$    0.193  &    0.318 $\pm$    0.087 &   7 &   7 &   7 &   0\\
  02212 & 13.50 & 0.15 &    3.568 $\pm$    0.085 &    0.342 $\pm$    0.111 &    1.095 $\pm$    0.142  &    0.547 $\pm$    0.131 &   7 &   8 &   8 &   0\\
  03122 & 14.65 & 0.15 &    4.010 $\pm$    1.237 &    0.258 $\pm$    0.199 &    1.400 $\pm$    0.389  &    0.414 $\pm$    0.172 &  23 &  23 &   0 &   0\\
  03288 & 15.50 & 0.15 &    2.832 $\pm$    1.100 &    0.139 $\pm$    0.127 &    1.400 $\pm$    0.488  &    0.222 $\pm$    0.197 &   0 &   6 &   0 &   0\\
  03552 & 13.00 & 0.15 &   26.656 $\pm$    9.734 &    0.016 $\pm$    0.009 &    1.000 $\pm$    0.462  &    0.025 $\pm$    0.011 &   0 &   0 &   4 &   0\\
  03554 & 15.94 & 0.15 &    1.660 $\pm$    0.524 &    0.261 $\pm$    0.190 &    1.400 $\pm$    0.477  &    0.417 $\pm$    0.197 &  15 &  15 &   0 &   0\\
  03753 & 15.13 & 0.15 &    2.071 $\pm$    0.106 &    0.365 $\pm$    0.082 &    1.531 $\pm$    0.108  &    0.414 $\pm$    0.113 &  13 &  15 &  13 &   0\\
  03753 & 15.13 & 0.15 &    1.803 $\pm$    0.447 &    0.352 $\pm$    0.250 &    1.530 $\pm$    0.366  &    0.398 $\pm$    0.213 &  18 &  20 &   0 &   0\\
  04055 & 14.90 & 0.15 &    2.204 $\pm$    0.078 &    0.415 $\pm$    0.071 &    1.400 $\pm$    0.000  &    0.644 $\pm$    0.138 &  49 &  51 &   0 &   0\\
  04401 & 15.50 & 0.15 &    1.801 $\pm$    0.535 &    0.343 $\pm$    0.204 &    1.400 $\pm$    0.441  &    0.549 $\pm$    0.259 &   0 &  13 &   0 &   0\\
\enddata
\end{deluxetable}

\end{document}